\documentstyle[preprint,aps,floats,epsfig]{revtex}
\draft
\begin{document}
\preprint{Phys. Rev. C (1999) in press.}
\title{Kaon differential flow in relativistic heavy-ion collisions}
\bigskip
\author{Bao-An Li\footnote{Email: Bali@navajo.astate.edu}$^{\rm a}$,
Bin Zhang\footnote{Email: Bzhang@kopc1.tamu.edu}$^{\rm b}$,
Andrew T. Sustich\footnote{Email: Sustich@navajo.astate.edu}$^{\rm a}$
and C. M. Ko\footnote{Email: Ko@cycomp.tamu.edu}$^{\rm b}$}
\address{$^{\rm a}$Department of Chemistry and Physics\\
Arkansas State University, P.O. Box 419\\
State University, Arkansas 72467-0419, USA}
\address{$^{\rm b}$Cyclotron Institute and Physics Department,\\
Texas A\&M University, College Station, Texas 77843, USA}
\maketitle

\begin{abstract}
Using a relativistic transport model, we study the azimuthal
momentum asymmetry of kaons with fixed transverse momentum, i.e.,
the differential flow, in heavy-ion collisions at beam momentum of
6 GeV/c per nucleon, available from the Alternating Gradient
Synchrotron (AGS) at the Brookhaven National Laboratory (BNL). We
find that in the absence of kaon potential the kaon differential
flow is positive and increases with transverse momentum as that of
nucleons. The repulsive kaon potential as predicted by theoretical
models, however, reduces the kaon differential flow, changing it to
negative for kaons with low transverse momenta. Cancellation
between the negative differential flow at low momenta and the
positive one at high momenta is then responsible for the
experimentally observed nearly vanishing in-plane transverse flow
of kaons in heavy ion experiments.
\end{abstract}
\noindent{PACS number(s): 25.75.Ld, 13.75.Jz, 21.65.+f}
\newpage

Since the work of Kaplan and Nelson \cite{kn} on the possibility of
a kaon condensation in the core of neutron stars, there have been
many theoretical studies of kaon properties in dense matter
\cite{kaon}. It is now generally agreed that a kaon has a weak
repulsive potential in nuclear matter while an antikaon has instead
a strong attractive one. The latter is responsible for the
existence of the kaon condensation that was originally proposed in
Ref. \cite{kn}. Brown and Bethe \cite{bb} have further argued that
because of the softening of the nuclear equation of state due to
the kaon condensation, the maximum mass of neutron stars would
decrease, hence allowing for the existence of mini black holes.

Since a hot dense matter can be created in the initial stage of
high energy heavy ion collisions, a unique opportunity thus exists
for studying the kaon in-medium properties
\cite{medium1,medium2,medium3,medium4}. Indeed, Li {\it et al.}
\cite{gqli1} have shown that it is possible to extract the
information on kaon potential or dispersion relation in dense
matter from the kaon collective flow in heavy ion collisions.
Specifically, they have found, based on the relativistic transport
model, that the repulsive kaon potential in nuclear medium would
reduce its flow relative to that of nucleons, while the flow of
antikaons would be similar to that of nucleons due to their
attractive potential \cite{kminus}. Since then, many theoretical
studies \cite{liko96,wang97,bra97,fuchs98,gqli98} have been carried
out, and all have reached the same conclusion. These theoretical
studies have stimulated a number of experiments at both GSI and
AGS. All experiments have shown that kaons have very small, if not
zero, flow, i.e., both the kaon average in-plane transverse
momentum as a function of rapidity and its slope at mid-rapidity
are consistent with zero within the experimental error bars
\cite{reis98}. For example, both the FOPI \cite{ritman,best} and the 
KaoS \cite{eos} collaboration have
found a negligible flow for $K^{+}$ in collisions of Ni+Ni at
$E_{\rm beam}/A$=1.93 GeV and Au+Au at 1 GeV, respectively. Several
collaborations at the AGS/BNL have also carried out the flow
analysis for $K^{+}$, $K^{-}$, and $K^0_s$ in Au+Au collisions at
beam energies from 2 to 12 GeV/nucleon, and no statistically
significant kaon transverse flow has been seen either
\cite{ogi97,ogi98,pol97,vol98}.

There are suggestions that the observed vanishing kaon flow may be
due to the fact that kaons are produced isotropically from
hadron-hadron collisions \cite{ogi98} or these colliding hadrons
have opposite transverse flows, leading thus to a reduced
collective flow of produced kaons \cite{dav98}. However, all
transport mode calculations
\cite{gqli1,liko96,wang97,bra97,fuchs98,gqli98} have shown that the
initial transverse flow of produced kaons is positive, similar to
that of nucleons, and the resulting vanishing kaon flow is due to
the repulsive kaon potential as first demonstrated by Li {\it et
al.} \cite{gqli1}.

Vanishing transverse flow has also been seen for nucleons in heavy
ion collisions at an incident energy called the balance energy.
Using the differential flow analysis, which measures the azimuthal
momentum asymmetry at fixed transverse momentum, two of present
authors \cite{lisu98} have shown that the zero transverse flow is
due to the cancellation between the positive flow of nucleons with
high transverse momentum and the negative flow of those with low
transverse momentum. The latter results from the effect of an
attractive nuclear mean-field potential. In this paper, we carry
out a similar differential flow analysis for kaons in order to
better understand the physics underlying the observed vanishing
kaon flow. We shall show that the repulsive kaon potential in
nuclear medium can indeed reduce the primordial positive in-plane
transverse flow of kaons to zero or even negative values, leading
to a strong differential flow, i.e., kaons with high and low
transverse momenta flow in opposite directions. A vanishing total
in-plane traverse flow is, nevertheless, obtained when the kaon
differential flow is integrated over the transverse momentum.

Our study is based on a relativistic transport model (ART1.0) for
heavy-ion collisions at AGS energies \cite{art}. In this model the
imaginary part of kaon self-energy is approximately treated by
scatterings of the kaon with other hadrons, and its real part is
given by the mean-field potential. We shall concentrate only on the
collective flow of $K^0$, so the Coulomb potential does not need to
be considered. Various approaches have been used to evaluate the
kaon dispersion relation in dense matter \cite{kn,kaon}. As an
illustration, we shall use in the present study the $K^+$
dispersion relation that is determined from the kaon-nucleon
scattering length $a_{KN}$ using the impulse approximation, i.e.,
\begin{equation}
\omega(p,\rho_b)=[m_K^2+p^2-4\pi (1+\frac{m_K}{m_N})a_{KN}\rho_b]^{1/2},
\end{equation}
where $m_K$ and $m_N$ are the kaon and nucleon masses,
respectively; $\rho_b$ is the baryon density and $a_{KN}\approx
-0.255$ fm is the isospin-averaged kaon-nucleon scattering
length. The $K^+(K^0)$ potential in nuclear medium can then be
defined as
\begin{equation}\label{pot}
U(p,\rho_b)=\omega(p,\rho_b)-(m_K^2+p^2)^{1/2},
\end{equation}
which gives a repulsive potential of about 30 MeV for a kaon at
rest in normal nuclear matter.

The transverse collective flow has usually been studied by
analyzing the average transverse momentum per particle in the
reaction plane as a function of rapidity $y$\cite{dani}, i.e.,
\begin{equation}\label{tflow}
< p_x/A >(y)=\frac{1}{A(y)}\sum_{i=1}^{A(y)} p_{ix}
=\frac{1}{A(y)}\int p_t \frac{dN}{dp_t}<\cos\phi>(y,p_t) dp_t,
\end{equation}
where $A(y)$ is the number of particles at rapidity $y$, $\phi$ is
the azimuthal angle with respect to the reaction plane, and
\begin{equation}\label{diff}
<\cos\phi>(y,p_t)\equiv \left(\frac{dN}{dp_t}\right)^{-1}
\int \cos\phi\frac{d^2N}{dp_td\phi}d\phi.
\end{equation}
A non-vanishing $<\cos\phi>(y,p_t)$ indicates the existence of an
azimuthally asymmetric transverse flow at the rapidity $y$ and
transverse momentum $p_t$. It is related to the first Fourier
coefficient in the expansion \cite{vol97,pos98},
\begin{equation}
\frac{d^2N}{dp_td\phi}=\frac{dN}{dp_t}
[1+\sum_{i=1}^{\infty}2v_i(y,p_t)\cos(i\phi)],
\end{equation}
i.e., $<\cos\phi>(y,p_t)=v_1(y,p_t)$. In Ref. \cite{lisu98}, the
$p_t$ dependence of $<\cos\phi>$ has been named the {\it
differential} flow. It has been found that the differential
analysis of nucleon flow at the balance energies is useful both for
understanding the disappearance of nuclear collective flow and for
extracting the information on the nuclear equation of state
\cite{lisu98}.

The transverse flow of baryons in heavy-ion collisions is an
experimentally well established fact at all energies. At beam
momentum 6 GeV/c, about half of the kaons are produced from
collisions between baryons, one thus expects that they would obtain
some collectivity as a result of the baryon collective flow.
Meson-baryon collisions almost account for the other half. Since
pions have a small directed flow, the directed flow of these kaons
is thus weaker than that of baryons. This is indeed seen in our
calculations, e.g., for a soft equation of
state, the average $p_x/m$ of those protons at $y/y_{\rm c.m.}=0.5$
is around $0.1$c and is about twice as large as that of kaons 
(kaon potential is not turned on).
Because of the large anisotropy of the baryon distribution,
especially in the later stage of the reaction, due to the baryon
transverse flow, a repulsive kaon potential would then cause kaons
to acquire a distribution that further differs from that of
baryons. To demonstrate the effect of kaon potential, we compare
results obtained with and without the kaon potential.

In Fig. 1, the average kaon transverse momentum (scaled by the kaon
mass) in the reaction plane is shown as a function of rapidity
(scaled by the beam rapidity) for Au+Au collisions at an impact
parameter of 4 fm and beam momentum of 6 GeV/c per nucleon.
Experimental data from this reaction are being analyzed by both the
E917 \cite{ogi98} and E895 \cite{e895} collaborations. The open
(filled) circles are the results obtained without (with) the kaon
mean-field potential. As one expects, in the absence of potential
kaons flow in the same direction as nucleons. However, there is
essential no flow in the whole rapidity range when the kaon
mean-field potential is included in the transport model. This
observation is in agreement with earlier findings at both lower
\cite{gqli1} and higher energies \cite{liko96}, i.e., the kaon
potential reduces its flow or even changes the flow direction. This
strong dependence of the kaon flow on its potential in nuclear
matter makes the kaon flow analysis a valuable tool to study the
kaon in-medium properties.

To examine more closely the origin of vanishing kaon flow in the
presence of a repulsive potential, we have carried out a
differential flow analysis for kaons. Since we are considering
collisions of symmetric nuclei, we include all kaons in the
rapidity range of ${\rm abs}(y/y_{\rm cm})\leq 0.5$ by flipping the
sign of $<\cos\phi>$ for $y\leq 0$ in Eq. (\ref{diff}), so an
improved statistics is obtained in evaluating the differential
flow. The results are shown in Fig. 2, and it is seen that
irrespective of the kaon transverse momentum its differential flow
is always positive when no potential is included. The value at
$p_t$ around 1 GeV/c is significantly below those at other
transverse momenta, and this is related to the fact that these
kaons are mainly produced when both the projectile and target
spectator matters are still present. With the repulsive kaon
potential given by Eq. (\ref{pot}), the differential flow of kaons
with transverse momenta less than about 0.8 GeV/c becomes negative
while that of kaons with higher transverse momenta remains
positive. This change in the dependence of kaon differential flow
on the transverse momentum follows from the fact that the force
acting on a kaon is inversely proportional to its energy, so low
energy kaons are more strongly repelled by baryons than high energy
ones. The latter thus remain to flow in the same direction as
baryons. Integrating over the transverse momentum distribution
leads to a cancellation between opposite flows at low and high
transverse momenta. The resulting kaon in-plane flow is therefore
small as shown in Fig. 1.

The differential kaon flow has recently been studied by the E877
collaboration \cite{pol97,vol98}, and it has indeed been found that
there is a negative (positive) flow for $K^+$ with low (high)
transverse momenta. A similar analysis for kaon differential flow
in the SIS/GSI experiments would be very useful in understanding
the observed vanishing kaon flow.

In summary, using the relativistic transport model (ART) for
heavy-ion collisions, we have studied the differential flow of
kaons. We have found that the experimentally observed nearly zero
in-plane transverse flow of kaons produced in heavy-ion collisions
is due to the cancellation between their negative and positive
differential flows at low and high transverse momenta. While the
positive differential flow reflects the strong positive baryonic
flow, the negative differential flow is caused by a repulsive
in-medium kaon potential. The differential flow analysis thus
provides a more detailed picture of kaon flow than the standard
transverse flow analysis, and is therefore very useful for probing
the in-medium kaon properties in heavy ion collisions.

This work was supported in part by NSF Grant No. PHY-9870038, the
Robert A Welch Foundation under Grant A-1358, and the Texas
Advanced Research Program.

\newpage
\begin{figure}
\centerline{\epsfig{file=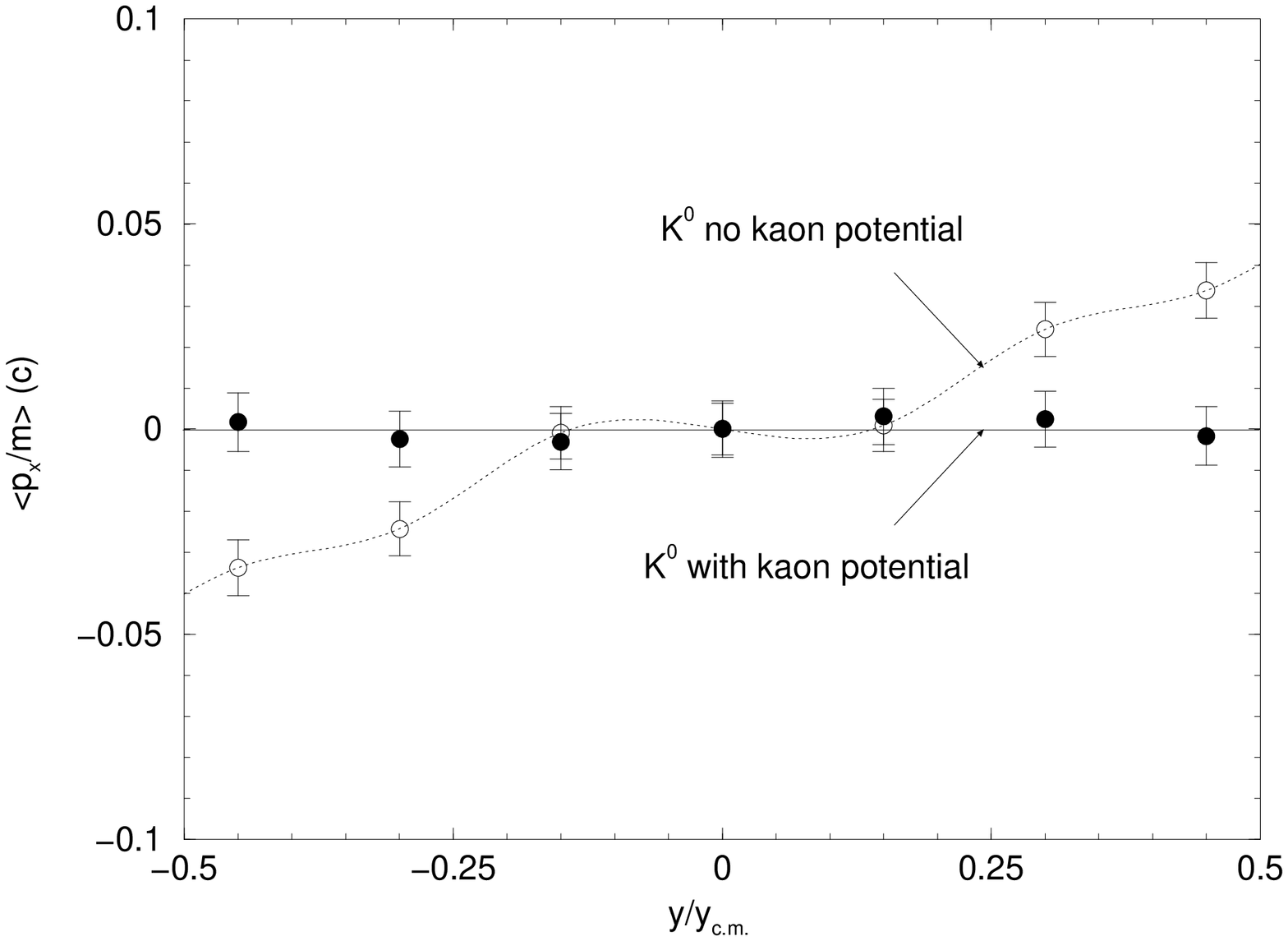,height=4in,width=3.8in,angle=-90}}
\vspace{0.5in}
\caption{The average transverse momentum of $K^0$
in the reaction plane for Au+Au reactions at $P_{\rm beam}/A=$ 6
GeV/c and an impact parameter of 4 fm. The open (filled) circles
are the results obtained without (with) the kaon mean-field
potential.}
\label{fig1}
\end{figure}

\newpage
\begin{figure}
\centerline{\epsfig{file=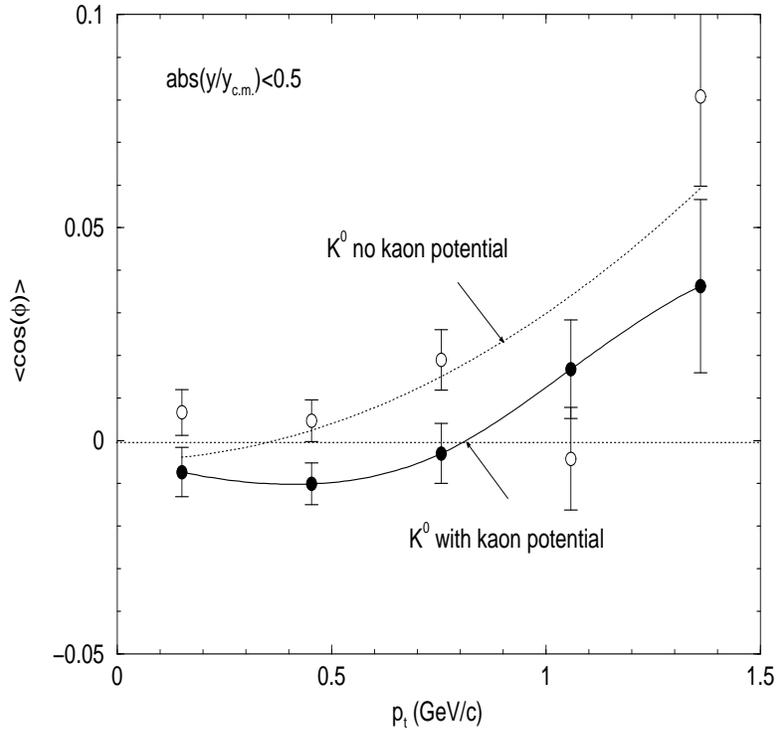,height=4in,width=3.8in,angle=-90}}
\vspace{0.5in}
\caption{The $K^0$ azimuthal asymmetry as a function of
transverse momentum for the same reaction as in Fig. 1. The open
(filled) circles are the results obtained without (with) the kaon
mean-field potential.}
\label{fig2}
\end{figure}

\end{document}